\documentclass{article}
\usepackage{slashed,amsmath,amssymb,enumitem}
\usepackage[papersize={8.5in,11in}]{geometry}
\begin{document}
\title{Ultraviolet Complete Quantum Gravity}
\author{J. W. Moffat\\~\\
Perimeter Institute for Theoretical Physics, Waterloo,\\ Ontario N2L 2Y5, Canada\\
and\\
Department of Physics and Astronomy, University of Waterloo,\\
Waterloo, Ontario N2L 3G1, Canada}
\date{\today}
\maketitle
\begin{abstract}%
An ultraviolet complete quantum gravity theory is formulated in which vertex functions in Feynman graphs are entire functions and the propagating graviton is described by a local, causal propagator. The cosmological constant problem is investigated in the context of the ultraviolet complete quantum gravity.

\end{abstract}



\section{Introduction}

As is well-known, the problem with perturbative quantum gravity based on a point-like graviton and a local field theory formalism is that the theory is not renormalizable~\cite{Veltman,Van,Van2}. Due to the Gauss-Bonnet theorem, it can be shown that the one-loop graviton calculation is renormalizable but two-loop is not~\cite{Sagnotti}. Moreover, gravity-matter interactions are not renormalizable at any loop order. In an ultraviolet (UV) complete electroweak (EW) theory~\cite{Moffat,Moffat2} the need for renormalization of the theory was removed and this allowed for the absence of a scalar degree of freedom and a Higgs particle in the action. This opens up the possibility of constructing viable UV complete theories that are not renormalizable, such as quantum gravity.

We develop an UV complete quantum gravity theory by extending the UV complete EW theory ~\cite{Moffat,Moffat2}. The quantum field theory (QFT) used to formulate our UV complete quantum gravity is based on having vertex functions associated with the coupling strengths described by {\it entire functions}. Important properties of entire functions are described in~\cite{Moffat}. A feature of the entire functions is that they do not possess poles in the finite complex plane and they thus do not correspond to physical particles in the particle spectrum. They will have poles and a possible essential singularity at infinity.  The coupling functions at the vertices in Feynman diagrams do not violate unitarity and satisfy the Cutkosky rules for the S-matrix scattering amplitudes~\cite{Cutkosky}. For quantum gravity the coupling constant $G$ is no longer the Newtonian constant $G_N$. The minimal choice $\sqrt{G}=\sqrt{G_N}$ leads to an unrenormalizable quantum gravity, whereas for $\sqrt{G}=\sqrt{G_N}{\cal F}$, where ${\cal F}$ is an entire function, the theory can be UV complete and satisfy unitarity to all orders of perturbation theory. The UV completeness is guaranteed by having entire functions at the vertices that vanish rapidly enough in Euclidean momentum space as $p^2\rightarrow\infty$.

In the entire vertex functions, constants $\Lambda_{\rm SM}$ and $\Lambda_G$ determine the energy scales in the standard model (SM) of particle interactions and gravitation, respectively. The SM energy scale $\Lambda_{SM}$ and the gravity energy scale $\Lambda_G$ are determined by the UV complete SM particle interactions as compared to the graviton interactions. The SM particle radiative corrections have an energy scale $\Lambda_{SM} > 1$ TeV or a length scale $\ell_{SM} < 10^{-16}$ cm, whereas the graviton radiative corrections remain local at a lower energy scale. Thus, the fundamental energy scales in the theory are determined by the underlying physical nature of the particles and fields and do not correspond to arbitrary cut-offs, which destroy the gauge invariance, unitarity and local Lorentz invariance of the field theory. The accurate values of the energy scales $\Lambda_{SM}$ and $\Lambda_G$ can be determined by experiment. {\it They are to be considered fundamental constants of nature}.
In our quantum gravity theory, the graviton-graviton and graviton-matter interactions are nonlocal due to the nonlocal entire functions at vertices in Feynman graphs. Unfortunately, there are no direct quantum gravity scattering amplitude data available to test the validity of our theory.

In Section 2, we describe the action of the theory and formulate the UV complete quantum gravity as a perturbative theory. In Section 3, we investigate the vacuum polarization in quantum gravity and gravity coupled to SM particles. The cosmological constant problem is examined in Section 4, and a possible solution is proposed. We end in Section 5, with concluding remarks.

\section{\bf UV Complete Quantum Gravity}

We begin with the four-dimensional action for gravity:
\begin{equation}
\label{Grav}
S=S_{\rm Grav}+S_M,
\end{equation}
where
\begin{equation}
\label{Gravaction}
S_{\rm Grav}=-\frac{1}{16\pi G_N}\int d^4x\sqrt{-g}(R+2\lambda),
\end{equation}
and $S_M$ is the matter action. Here, we use the notation: $\mu,\nu=0,1,2,3$, $g={\rm det}(g_{\mu\nu})$, the
metric signature of Minkowski spacetime is $\eta_{\mu\nu}={\rm diag}(+1,-1,-1,-1)$, $G_N$ is Newton's gravitational constant and $\lambda$ is the cosmological constant. The Ricci tensor is defined by
\begin{equation}
R_{\mu\nu}=\partial_\alpha{\Gamma^\alpha}_{\mu\nu}-\partial_\nu{\Gamma^\alpha}_{\mu\alpha}+{\Gamma^\alpha}_{\mu\nu}{\Gamma^\beta}_{\alpha\beta}
-{\Gamma^\alpha}_{\mu\beta}{\Gamma^\beta}_{\alpha\nu}.
\end{equation}

We assume that the matter action $S_M$ is modified to incorporate a nonlocal coupling to gravity. As usual, the generalized energy momentum tensor $S_{\mu\nu}$ is given by
\begin{equation}
-\frac{2}{\sqrt{-g}}\frac{\delta S_M}{\delta g^{\mu\nu}}=S_{\mu\nu}.
\end{equation}
Here, we have assumed that
\begin{equation}
S_{\mu\nu}={\cal F}^2(\Box/2\Lambda_G^2)T_{\mu\nu},
\end{equation}
where ${\cal F}$ is an entire function and
\begin{equation}
\Box=g^{\mu\nu}\nabla_\mu\nabla_\nu=\frac{1}{\sqrt{-g}}(\partial_\mu\sqrt{-g}g^{\mu\nu}\partial_\nu),
\end{equation}
and $T_{\mu\nu}$ is the standard stress-energy-momentum tensor in the absence of nonlocal coupling.  Moreover, $\Lambda_G$ is a fundamental gravitational energy scale. The generalized Einstein field equations are given by
\begin{equation}
\label{Einsteineq}
G_{\mu\nu}=8\pi G_NS_{\mu\nu},
\end{equation}
where $G_{\mu\nu}=R_{\mu\nu}-\frac{1}{2}g_{\mu\nu}R$. Since ${\cal F}$ is an invertible function, we can rewrite the field equations in the form:
\begin{equation}
{\cal F}^{-2}(\Box/2\Lambda_G^2)G_{\mu\nu}=8\pi G_NT_{\mu\nu}.
\end{equation}
We note that ${\cal F}^{-2}$ is a differential operator acting on the Einstein tensor $G_{\mu\nu}$. The field equations in vacuum for $T_{\mu\nu}=0$ are given by
\begin{equation}
{\cal F}^{-2}(\Box/2\Lambda_G^2)G_{\mu\nu}=0.
\end{equation}
Because of the Bianchi identities: $\nabla_\nu G^{\mu\nu}=0$ the field equations (\ref{Einsteineq}) yield the conservation law:
\begin{equation}
\nabla_\nu S^{\mu\nu}=0.
\end{equation}

We note that the Einstein-Hilbert action (\ref{Gravaction}) is not modified. It is the matter action $S_M$ that is modified by the nonlocal differential operator ${\cal F}^2$ to produce the smeared energy-momentum tensor $S_{\mu\nu}$. The choice between $T_{\mu\nu}=0$ and $S_{\mu\nu}=0$ will produce two different possible vacuum states.

The entire function ${\cal F}$ is analytic (holomorphic) in the complex $z$ plane, except for singularities at infinity. This means that we can expand ${\cal F}(\Box/2\Lambda^2_G)$ in a power series in $\Box/2\Lambda_G^2$.  The series expansion of ${\cal F}(\Box/2\Lambda_G^2)$ is not to be truncated, for this would induce instabilities and ghosts violating unitarity. We require that in the limit of classical gravity $S_{\mu\nu}$ reduces to $T_{\mu\nu}$.

In~\cite{Moffat} the vertex factors in the QFT formalism were described by {\it entire functions} in spacetime and in the Fourier transformed momentum space. These functions possess no poles in the finite complex momentum plane, so they {\it do not describe physical fields and do not violate unitarity}. We extend this QFT formalism to quantum gravity by postulating that the gravitational coupling is given by
\begin{equation}
\label{Ffunction}
\sqrt{G}(x)=\sqrt{G_N}{\cal F}\biggl(\Box(x)/2\Lambda_G^2\biggr).
\end{equation}
The Fourier transform of (\ref{Ffunction}) in a flat Euclidean momentum space is
\begin{equation}
\sqrt{G}(p^2)=\sqrt{G_N}{\cal F}\biggl(-p^2/2\Lambda_G^2\biggr).
\end{equation}
We require that on-shell: $G(0)=G_N$.

Consider the operator ${\cal F}(t)$ in momentum space represented as an infinite series in powers of $t=-p^2/2\Lambda_G^2$:
\begin{equation}
{\cal F}(t)=\sum_{j=0}^\infty\biggl[\frac{|c_j^2|}{(2j)!}\biggr]^{1/2}t^j.
\end{equation}
The function ${\cal F}(t)$ is an {\it entire} function of $t$, which is analytic (holomorphic) in the finite complex momentum $t$-plane. Thus, ${\cal F}$ has no singularities in the finite complex $t$-plane. However, it will have a pole or an essential singularity at infinity, otherwise, by Liouville's theorem it is constant. This avoids non-physical singularities occurring in scattering amplitudes which violate the S-matrix unitarity and the Cutkosky rules~\cite{Cutkosky}. We can distinguish three cases~\cite{Efimov,Efimov2,Efimov3}:
\begin{enumerate}
\item  \quad\quad ${\rm lim\,sup}_{j\rightarrow\infty}\vert c_j\vert^{\frac{1}{j}}=0,$
\item  \quad\quad ${\rm lim\,sup}_{j\rightarrow\infty}\vert c_j\vert^{\frac{1}{j}}={\rm const. }< \infty,$
\item  \quad\quad ${\rm lim\,sup}_{j\rightarrow\infty}\vert c_j\vert^{\frac{1}{j}}=\infty.$
\end{enumerate}
Let us consider the order of the entire functions ${\cal F}(t)$. For (1) the order is $\gamma < \frac{1}{2}$:
\begin{equation}
\vert{\cal F}(t)\vert < \exp(\alpha\vert t\vert^\gamma),
\end{equation}
where $\alpha > 0$. An entire function with this property is
\begin{equation}
{\cal F}(z)=\sum^\infty_{n=0}\frac{z^n}{\Gamma(bn+1)}\quad b > 2,
\end{equation}
where $\Gamma$ is the gamma function. For type (1) entire functions, we know that ${\cal F}(t)$ does not decrease along any direction in the complex $z$ plane. Therefore, we cannot use this type of function to describe our coupling function $G(t)=G_N{\cal F}^2(t)$, for it will not produce a UV finite perturbation theory.

For case (2), the entire functions ${\cal F}(t)$ are of order $\gamma=1/2$ and we have
\begin{equation}
\vert{\cal F}(t)\vert \leq \exp(\alpha\sqrt{|t|}).
\end{equation}
This type of entire function {\it can decrease along one direction} in the complex $t$-plane.

In the case (3), we have $\gamma > 1/2$ and now
\begin{equation}
\vert{\cal F}(t)\vert \leq \exp(f(t)\vert t\vert),
\end{equation}
where $f(\vert t\vert)$ is a positive function which obeys the condition $f(\vert t\vert) > \alpha\vert t\vert^{1/2}$ as $\vert t\vert\rightarrow \infty$ for any $\alpha > 0$. These functions can decrease along whole regions for $\vert t\vert\rightarrow \infty$ and can be chosen to describe the coupling functions and lead to a UV finite, unitary perturbation theory. We note that a consequence of the fundamental theorem of algebra is that ``genuinely different" entire functions {\it cannot dominate each other}, i.e., if $f$ and $g$ are entire functions and $|f| \leq |g|$ everywhere, then $f=ag$ for some complex number $a$. This theorem plays an important role in providing a uniqueness of choice of entire functions ${\cal F}(-p^2/\Lambda_G^2)$ for $p^2\rightarrow\infty$.

The quantization proceeds from splitting the free and interacting pieces of the Lagrangian density, and replacing vertex functions by entire functions. The Lagrangian is gauge invariant to all orders, guaranteeing that there is no coupling to unphysical quanta, such as longitudinal gravitons.

The path integral formalism for quantization can be performed using the definition
\begin{equation}
\langle 0\vert T^*(O[f])\vert 0\rangle_{\cal F}=\int[Df]\mu[f]({\rm gauge\, fixing})O[f]\exp(i\hat S[f]),
\end{equation}
where $f$ is a generic field operator and ${\hat S}$ is the regulated action. On the left-hand side we have the regulated vacuum expectation value of the $T^*$-ordered product of an arbitrary operator $O[f]$ formed from the local fields $f$. The subscript ${\cal F}$ signifies that a Lorentz distribution has been used constructed from entire functions maintaining unitarity. Moreover, $\mu[f]$ is a gauge invariant measure factor and there is a gauge fixing factor, both of which are needed to maintain perturbative unitarity in gauge theories.

The Feynman rules are obtained as follows: Every leg of a diagram is connected to a local propagator,
\begin{equation}
\label{regpropagator}
D(p^2)=\frac{i}{p^2-M^2+i\epsilon}
\end{equation}
and every vertex has a factor ${\cal F}(p^2)$, where $p$ is the momentum attached to the propagator $D(p^2)$. The formalism is set up in Minkowski spacetime and loop integrals are formally defined in Euclidean space by performing a Wick rotation. The Wick rotation can be conducted for the entire function ${\cal F}(p^2)$, {\it thereby allowing for an analytic continuation from Euclidean space to Minkowski space}, provided ${\cal F}(p^2)$ vanishes in the limit $p^2\rightarrow\pm\infty$ for both space-like and time-like values of $p^2$. A possible choice of an entire function that satisfies the requirement of vanishing for space-like and time-like $p^2\rightarrow\pm\infty$ is: ${\cal F}(p^2)=\exp(-\vert p\vert^2/\Lambda_G^2)$ where $\vert p\vert=\vert(p_\mu p^\mu)\vert^{1/2}$. The Lorentz distribution function ${\cal F}$ must be chosen to perform an explicit calculation in perturbation theory. For a general entire function ${\cal F}$, we can perform a regularization ${\cal F}R^\delta$, such that a Wick rotation $p_0=ip_4$ can be carried out for a chosen $R^\delta$ that vanishes in the upper-half complex plane, allowing an analytic continuation to Euclidean momentum space with the subsequent limit: $\delta\rightarrow 0$~\cite{Efimov,Moffat}. We do not know the unique choice of ${\cal F}$. However, once a choice for the function is made, then the theory and the perturbative calculations are uniquely fixed. All loops contain at least one vertex function ${\cal F}$ and therefore are ultraviolet finite.

We shall now examine the gravitational sector in more detail. Our quantum gravity theory can be formulated as a perturbative theory by expanding around any fixed, classical metric background~\cite{Veltman}:
\begin{equation}
\label{background}
g_{\mu\nu}={\bar g}_{\mu\nu}+h_{\mu\nu},
\end{equation}
where ${\bar g}_{\mu\nu}$ is any smooth background metric field, e.g., a flat Minkowski spacetime or a de Sitter spacetime. We maintain local gauge invariance of the gravitational calculations under the gauge group of the fixed background metric, e.g., for a fixed Minkowski metric background the action is invariant under local Poincar\'{e} transformations, while for a de Sitter background metric the action will be invariant under the group of de Sitter transformations. We argue that although we lose general covariance in our perturbation calculations of gravitational scattering amplitudes, the basic physical properties such as finiteness of loop amplitudes, gauge invariance and unitarity will ultimately lead to correct and reliable physical conclusions for $E\lesssim M_{\rm PL}$ where $M_{\rm PL}\sim 10^{19}$ GeV is the Planck energy.

For the sake of simplicity, we only consider expansions about flat Minkowski spacetime. If we wish to include the cosmological constant, then we cannot strictly speaking expand about flat spacetime. This is to be expected, because the cosmological constant produces a curved spacetime even when the energy-momentum tensors $S_{\mu\nu}=0$ and $T_{\mu\nu}=0$. Therefore, we should use the expansion (\ref{background}) with a curved background metric. But for energy scales encountered in particle physics, the curvature is very small, so we can approximate the perturbation calculation by using the flat spacetime expansion and trust that the results are valid in general for curved spacetime
backgrounds, including the cosmological constant.

Let us define ${\bf g}^{\mu\nu}=\sqrt{-g}g^{\mu\nu}$ and ${\bf g}=
{\rm det}({\bf g}^{\mu\nu})$. It can be shown that $\sqrt{-g}=\sqrt{-{\bf g}}$
and $\partial_\rho{\bf g}={\bf g}_{\alpha\beta}\partial_\rho{\bf g}^{\alpha\beta}{\bf g}$. We can write the gravitational Lagrangian density ${\cal L}_{\rm grav}$ in the first-order form~\cite{Goldberg}:
\begin{align}
\label{action}
{\cal L}_{\rm grav}=\frac{1}{2\kappa^2}\biggl({\bf g}^{\rho\sigma}{\bf g}_{\lambda\mu} {\bf g}_{\kappa\nu}
-\frac{1}{2}{\bf g}^{\rho\sigma} {\bf g}_{\mu\kappa}{\bf g}_{\lambda\nu}
-2\delta^\sigma_\kappa\delta^\rho_\lambda{\bf
g}_{\mu\nu}\biggr)\partial_\rho{\bf g}^{\mu\kappa} \partial_\sigma{\bf
g}^{\lambda\nu}\nonumber\\
-\frac{1}{\alpha\kappa^2}\partial_\mu{\bf g}^{\mu\nu}\partial_\kappa{\bf g}^{\kappa\lambda}{\bf g}_{\nu\lambda}
+{\bar C}_\nu\partial_\mu {X^{\mu\nu}}_\lambda C^\lambda,
\end{align}
where $\kappa^2=32\pi G_N$. We have added a gauge fixing term with the parameter $\alpha$, $C_\mu$ is the Fadeev-Popov ghost field and ${X^{\mu\nu}}_\lambda$ is a differential operator.

We expand the local interpolating graviton field ${\bf g}^{\mu\nu}$ as
\begin{equation}
\label{weakg}
{\bf g}^{\mu\nu}=\eta^{\mu\nu}+\kappa\gamma^{\mu\nu}.
\end{equation} Then,
\begin{equation}
\label{upperweakg}
{\bf g}_{\mu\nu}=\eta_{\mu\nu}-\kappa\gamma_{\mu\nu}
+\kappa^2{\gamma_\mu}^\alpha{\gamma_\alpha}_\nu-\kappa^3\gamma_{\mu\alpha}{\gamma^\alpha}_\beta{\gamma^\beta}_\nu+O(\kappa^4).
\end{equation}
We also have
\begin{equation}
\label{detg}
\frac{1}{\sqrt{-{\bf g}}}=1-\frac{\kappa}{2}\gamma+O(\kappa^2),
\end{equation}
where $\gamma={\gamma^\lambda}_\lambda$. We note that ${\cal F}$ is a function of the covariant derivative operator $\Box=\nabla_\mu\nabla^\mu$.

The local gravitational Lagrangian density is expanded as
\begin{equation}
{\cal L}_{\rm grav}={\cal L}^{(0)}+\kappa{\cal L}^{(1)}
+\kappa^2{\cal L}^{(2)}+....
\end{equation}
We obtain
\begin{align}
{\cal L}^{(0)}=\frac{1}{2}\partial_\sigma\gamma_{\lambda\rho}
\partial^\sigma\gamma^{\lambda\rho}
-\partial_\lambda\gamma^{\rho\kappa}
\partial_\kappa\gamma^\lambda_\rho
-\frac{1}{4}\partial_\rho\gamma\partial^\rho\gamma
-\frac{1}{\alpha}\partial_\rho\gamma^\rho_\lambda\partial_\kappa
\gamma^{\kappa\lambda}
+{\bar C}^\lambda\Box C_\lambda,
\end{align}
\begin{align}
{\cal L}^{(1)}
=\frac{1}{4}(-4\gamma_{\lambda\mu}\partial^\rho\gamma^{\mu\kappa}
\partial_\rho\gamma^\lambda_\kappa+2\gamma_{\mu\kappa}
\partial^\rho\gamma^{\mu\kappa}\partial_\rho\gamma
+2\gamma^{\rho\sigma}\partial_\rho\gamma_{\lambda\nu}
\partial_\sigma\gamma^{\lambda\nu}
-\gamma^{\rho\sigma}\partial_\rho\gamma\partial_\sigma\gamma
+4\gamma_{\mu\nu}\partial_\lambda\gamma^{\mu\kappa}
\partial_\kappa\gamma^{\nu\lambda})\nonumber\\
+{\bar C}^\nu\gamma_{\kappa\mu}\partial^\kappa\partial^\mu C_\nu
+{\bar C}^\nu\partial^\mu\gamma_{\kappa\mu}\partial^\kappa C_\nu
-{\bar C}^\nu\partial^\lambda\partial^\mu\gamma_{\mu\nu}C_\lambda
-{\bar C}^\nu\partial^\mu\gamma_{\mu\nu}\partial^\lambda
C_\lambda,
\end{align}
\begin{align}
{\cal L}^{(2)}=\frac{1}{4}[4\gamma_{\kappa\alpha}
\gamma^{\alpha\nu}
\partial^\rho\gamma^{\lambda\kappa}\partial_\rho\gamma_{\nu\lambda}
+(2\gamma_{\lambda\mu}\gamma_{\kappa\nu}-\gamma_{\mu\kappa}\gamma_{\nu\lambda})
\partial^\rho\gamma^{\mu\kappa}\partial_\rho\gamma^{\nu\lambda}\nonumber\\
-2\gamma_{\lambda\alpha}\gamma^\alpha_\nu\partial^\rho\gamma^{\lambda\nu}
\partial_\rho\gamma-2\gamma^{\rho\sigma}\gamma^\kappa_\nu
\partial_\rho\gamma_{\lambda\kappa}\partial_\sigma\gamma^{\nu\lambda}\nonumber\\
+\gamma^{\rho\sigma}\gamma^{\nu\lambda}\partial_\sigma\gamma_{\nu\lambda}
\partial_\rho\gamma-2\gamma_{\mu\alpha}\gamma^{\alpha\nu}
\partial^\lambda\gamma^{\mu\kappa}\partial_\kappa\gamma_{\nu\lambda}].
\end{align}

Let us consider the effects of an infinitesimal gauge transformation: $x^{'\mu}=x^\mu+\xi^\mu$, where $\xi^\mu$ is an infinitesimal vector quantity. We get
\begin{equation}
\delta{\bf g}^{\mu\nu}=-\xi^\lambda\partial_\lambda{\bf g}^{\mu\nu}+\partial_\rho\xi^\mu{\bf g}^{\rho\nu}+\partial_\rho\xi^\nu{\bf g}^{\mu\rho}-\partial_\alpha\xi^\alpha{\bf g}^{\mu\nu}.
\end{equation}
In the limit the gauge parameter $\alpha\rightarrow\infty$, the Lagrangian density ${\cal L}_{\rm grav}$ is invariant under the gauge transformation:
\begin{equation}
\delta\gamma^{\mu\nu}={X^{\mu\nu}}_\lambda\xi^\lambda,
\end{equation}
where ${X^{\mu\nu}}_\lambda$ is a differential operator:
\begin{equation}
{X^{\mu\nu}}_\lambda=-\partial_\lambda\gamma^{\mu\nu}
+\gamma^{\rho\nu}\delta_\lambda^\mu\partial_\rho
+\gamma^{\mu\sigma}\delta^\nu_\lambda\partial_\sigma-\gamma^{\mu\nu}\delta^\sigma_\lambda\partial_\sigma
+\frac{1}{\kappa}(\delta^\mu_\lambda\partial^\nu+\delta^\nu_\lambda\partial^\mu-\eta^{\mu\nu}\partial_\lambda).
\end{equation}
For the quantized theory, it is more useful for the gauge fixed theory to require BRST symmetry~\cite{Becchi,Tyutin}. We choose $\xi^\lambda=C^\lambda\sigma$, where $\sigma$ is a global anticommuting scalar. Then, the BRST transformation is
\begin{equation}
\delta\gamma^{\mu\nu}={X^{\mu\nu}}_\lambda C^\lambda\sigma,
\quad \delta {\bar C}^\nu=-\partial_\mu\gamma^{\mu\nu}
\biggl(\frac{2\sigma}{\alpha}\biggr),\quad \delta C_\nu=\kappa C^\mu \partial_\mu C_\nu\sigma.
\end{equation}

To make the nonlocal Lagrangian density gauge invariant, we must extend the local gauge transformations to nonlocal gauge transformations~\cite{Evens}. The nonlocal regularized Lagrangian density is invariant under the extended BRST transformations up to, but not including, $\kappa^2$~\cite{Cornish,Cornish2}:
\begin{equation}
\delta_0\gamma^{\mu\nu}={X^{(0)\mu\nu}}_\lambda
C^\lambda\sigma=\frac{1}{\kappa}(\partial^\nu C^\mu+\partial^\mu
C^\nu-\eta^{\mu\nu}\partial_\lambda C^\lambda)\sigma,
\end{equation}
\begin{equation}
\delta_1\gamma^{\mu\nu}={\cal F}^2{X^{(1)\mu\nu}}_\lambda C^\lambda\sigma
={\cal F}^2(-\partial_\lambda\gamma^{\mu\nu}C^\lambda+\gamma^{\rho\nu}\partial_\rho C^\mu
+\gamma^{\mu\sigma}\partial_\sigma C^\nu-\gamma^{\mu\nu}\partial_\lambda C^\lambda)\sigma,
\end{equation}
\begin{equation}
\delta_0{\bar C}^\nu=2\partial_\mu\gamma^{\mu\nu}\sigma,
\end{equation}
\begin{equation}
\delta_1C_\nu=\kappa{\cal F}^2C^\mu \partial_\mu C_\nu\sigma.
\end{equation}
The nonlocal gauge transformations can be derived that include $\delta_2$ to order $\kappa^2$ and those in higher orders in $\kappa$ ~\cite{Cornish,Cornish2}. In the present approach the vertices in Feynman graphs are entire functions, corresponding to nonlocal, scalar differential operators ${\cal F}$ in spacetime and scalar differential operators in momentum space. They do not describe physical propagating fields.

The local graviton propagator in the fixed gauge $\alpha=-1$ in momentum space is given by
\begin{equation}
D^G_{\mu\nu\rho\sigma}(p)
=\frac{\eta_{\mu\rho}\eta_{\nu\sigma}+\eta_{\mu\sigma}\eta_{\nu\rho}
-\eta_{\mu\nu}\eta_{\rho\sigma}}{p^2+i\epsilon},
\end{equation}
while the local graviton ghost propagator in momentum space is
\begin{equation}
D^{\rm ghost}_{\mu\nu}(p)=\frac{\eta_{\mu\nu}}{p^2+i\epsilon}.
\end{equation}

Because we have extended the gauge symmetry to nonlinear transformations, we must also supplement the
quantization procedure with an invariant measure
\begin{equation}
{\cal M}=\Delta({\bf g}, {\bar C}, C)D[{\bf g}_{\mu\nu}]D[{\bar C}_\lambda]D[C_\sigma]
\end{equation}
such that $\delta {\cal M}=0$.

We quantize by means of the path integral operation
\begin{equation}
\langle 0\vert T^*(O[{\bf g}])\vert 0\rangle_{\cal F}=\int[D{\bf g}]
\mu[{\bf g}]({\rm gauge\, fixing})O[{\bf g}]\exp(i\hat S_{\rm grav}[{\bf g}]).
\end{equation}
The quantization is carried out in the functional formalism by finding a measure factor
$\mu[{\bf g}]$ to make $[D{\bf g}]$ invariant under the classical symmetry. To ensure a correct gauge fixing scheme, we write
${\hat S}_{\rm grav}[{\bf g}]$ in the BRST invariant form with ghost fields; the ghost structure arises from exponentiating the Faddeev-Popov determinant~\cite{Fradkin}. The algebra of extended gauge symmetries is not expected to close off-shell, so one needs to introduce higher ghost terms (beyond the normal ones) into both the action and the BRST transformation. The BRST action will be regularized directly to ensure that all the corrections to the measure factor are included.

\section{Gravitational Vacuum Polarization}

A feature of our UV complete field theory is that the vertex function ${\cal F}(p^2)$ in momentum space is determined by the choice of entire functions. For a SM gauge boson, such as the $W$ or $Z$ boson connected to a standard model particle and anti-particle, the vertex function in Euclidean momentum space is
\begin{equation}
{\cal E}^{\rm SM}(p^2)={\cal E}\biggl(p^2/2\Lambda^2_{SM}\biggr),
\end{equation}
where $\Lambda_{SM}$ is the energy scale for the standard particle model. For a vertex with a graviton attached to a standard model particle and an
anti-particle, the vertex function will be
\begin{equation}
{\cal F}^G(p^2)={\cal F}\biggl(p^2/2\Lambda^2_G\biggr).
\end{equation}
Thus, when two vertices are drawn together to make a loop graph, the energy scales
$\Lambda_{SM}$ and $\Lambda_G$ will be determined by the external legs attached to the loop.
If we ignore the weak effects of gravity in SM calculations, then the graviton scale
$\Lambda_G$ can be ignored, as is usually the case in SM calculations.

The lowest order contributions to the graviton self-energy will include the graviton loops, the ghost field loop contribution and the measure loop contribution. In the perturbative quantum gravity theory the first order vacuum polarization tensor $\Pi^{\mu\nu\rho\sigma}$ must satisfy the
Slavnov-Ward identities~\cite{Medrano}:
\begin{equation}
\label{SlavnovWard}
p_\mu p_\rho D^{G\mu\nu\alpha\beta}(p)\Pi_{\alpha\beta\gamma\delta}(p)
D^{G\gamma\delta\rho\sigma}(p)=0.
\end{equation}
By symmetry and Lorentz invariance, the vacuum polarization tensor must have the form
\begin{align}
\Pi_{\alpha\beta\gamma\delta}(p)
=\Pi_1(p^2)p^4\eta_{\alpha\beta}\eta_{\gamma\delta}+\Pi_2
(p^2)p^4(\eta_{\alpha\gamma}\eta_{\beta\delta}
+\eta_{\alpha\delta}\eta_{\beta\gamma})\nonumber\\
+\Pi_3(p^2)p^2(\eta_{\alpha\beta}p_\gamma
p_\delta+\eta_{\gamma\delta}p_\alpha p_\beta)
+\Pi_4(p^2)p^2(\eta_{\alpha\gamma}p_\beta
p_\delta+\eta_{\alpha\delta}p_\beta p_\gamma\nonumber\\
+\eta_{\beta\gamma}p_\alpha p_\delta+\eta_{\beta\delta}p_\alpha
p_\gamma)+\Pi_5(p^2)p_\alpha p_\beta p_\gamma p_\delta.
\end{align}
The Slavnov-Ward identities impose the restrictions
\begin{equation}
\Pi_2+\Pi_4=0,\quad 4(\Pi_1+\Pi_2-\Pi_3)+\Pi_5=0.
\end{equation}

The basic lowest order graviton self-energy diagram is determined by~\cite{Leibbrandt,Brown}:
\begin{align}
\label{basic}
\Pi^{\rm basic}_{\mu\nu\rho\sigma}(p)=\frac{1}{2}\kappa^2\int
d^4q {\cal U}_{\mu\nu\alpha\beta\gamma\delta}(p,-q,q-p){\cal F}(q^2)
D^{G\alpha\beta\kappa\lambda}(q)\nonumber\\
\times D^{G\gamma\delta\tau\xi}(q-p){\cal
U}_{\kappa\lambda\tau\xi\rho\sigma}(q,p-q,-p){\cal F}((q-p)^2),
\end{align}
where ${\cal U}$ is the three-graviton vertex function:
\begin{align}
{\cal U}_{\mu\nu\rho\sigma\delta\tau}(q_1,q_2,q_3) =
-\frac{1}{2}[q_{2(\mu}q_{3\nu)}\biggl(2\eta_{\rho(\delta}\eta_{\tau)\sigma}
-\eta_{\rho\sigma}\eta_{\delta\tau}\biggr)\nonumber\\
+q_{1(\rho}q_{3\sigma)}\biggl(2\eta_{\mu(\delta}\eta_{\tau)\nu}
-\eta_{\mu\nu}\eta_{\delta\tau}\biggr)+\ldots],
\end{align}
where $A_{(\alpha}B_{\beta)}=\frac{1}{2}(A_\alpha B_\beta+A_\beta B_\alpha)$, and the ellipsis denotes similar contributions. The ghost-ghost graviton vertex operator is given by
\begin{equation}
{\cal W}_{\alpha\beta\lambda\mu}(q_1,q_2,q_3)=-\eta_{\lambda(\alpha}q_{1\beta)}q_{2\mu}+\eta_{\lambda\mu}q_{2(\alpha}q_{3\beta)}.
\end{equation}

To the basic self-energy diagram (\ref{basic}), we must add the ghost particle diagram contribution $\Pi^{\rm ghost}$ and the measure diagram contribution $\Pi^{\rm meas}$. The dominant finite contribution to the graviton self-energy will be of the form
\begin{equation}
\label{gravpol}
\Pi_{\mu\nu\rho\sigma}(p)\sim\kappa^2Q_{\mu\nu\rho\sigma}(\Lambda_G,p^2),
\end{equation}
where $Q_{\mu\nu\rho\sigma}(\Lambda_G,p^2)$ is a finite contribution. Renormalizability is no longer an issue for obtaining finite scattering amplitudes. The function ${{Q_{\mu}}^{\mu\sigma}}_\sigma(\Lambda_G,p^2)$ vanishes as $p^2\rightarrow 0$. Therefore, ${{{\Pi_\mu}^\mu}^\sigma}_\sigma(p)$ vanishes at $p^2=0$, as it should from gauge invariance to this order and for massless gravitons.

Let us now consider the dominant contributions to the vacuum density arising from the graviton loop corrections for which the loops consist of SM particles. As explained above, we perform the calculations by expanding about flat spacetime and trust that the results still hold for
an expansion about a curved metric background field, which is strictly required for a non-zero cosmological constant. Since
the scales involved in the final answer correspond to a very small curvature of spacetime for $E \ll M_{PL}$, we expect that our approximation is justified.

We adopt a model consisting of a photon field $A_\mu$ coupled to gravity.  We have for the vector field Lagrangian density~\cite{Duff}:
\begin{equation}
\label{LA}
{\cal L}_A=-\frac{1}{4}(-{\bf g})^{-1/2}{\bf g}^{\mu\nu}{\bf g}^{\alpha\beta}F_{\mu\alpha}F_{\nu\beta},
\end{equation}
where
\begin{equation}
F_{\mu\nu}=\partial_\nu A_\mu-\partial_\mu A_\nu.
\end{equation}

The Lagrangian density (\ref{LA}) is now expanded according to (\ref{weakg}), (\ref{upperweakg}) and (\ref{detg}). We find that
\begin{equation}
{\cal L}_A^{(0)}=-\frac{1}{4}\eta^{\mu\nu}\eta^{\alpha\beta}F_{\mu\alpha}F_{\nu\beta},
\end{equation}
and
\begin{equation}
{\cal L}_A^{(1)}=-\frac{1}{4}\biggl(\eta^{\mu\nu}\gamma^{\alpha\beta}+\eta^{\alpha\beta}\gamma^{\mu\nu}
-\frac{1}{2}\eta^{\mu\nu}\eta^{\alpha\beta}\gamma\biggr)F_{\mu\alpha}F_{\nu\beta}.
\end{equation}
We include in the Lagrangian density ${\cal L}_A^{(0)}$ an additional gauge-fixing piece $-\frac{1}{2}(\partial^\mu A_\mu)^2$. For a particular gauge no Faddeev-Popov ghost particles and diagrams contribute to the lowest order photon-graviton self-energy calculation~\cite{Duff}. The local photon propagator has the form
\begin{equation}
D^{\rm A}_{\mu\nu}(p)=\frac{\eta_{\mu\nu}}{p^2+i\epsilon}.
\end{equation}

The graviton-A-A vertex in momentum space is given by
\begin{align}
{\cal V}_{\alpha\beta\lambda\sigma}(q_1,q_2)
=\eta_{\lambda\sigma}
q_{1(\alpha}q_{2\beta)}-\eta_{\sigma(\beta}q_{1\alpha)}q_{2\lambda}
-\eta_{\lambda(\alpha}q_{1_\sigma}q_{2\beta)}\nonumber\\
+\eta_{\sigma(\beta}\eta_{\alpha)\lambda}q_1{\cdot q_2}
-\frac{1}{2}\eta_{\alpha\beta}(\eta_{\lambda\sigma}
q_1{\cdot q_2}-q_{1\sigma}q_{2\lambda}),
\end{align}
where $q_1,q_2$ denote the momenta of the two $As$ connected to the graviton with momentum $p$.

The lowest order correction to the graviton vacuum loop will have the form
\begin{align}
\label{PolV}
\Pi^{\rm GA}_{\mu\nu\rho\sigma}(p)
=-\kappa^2\int d^4q
{\cal V}_{\mu\nu\lambda\alpha}(p,q){\cal F}(q^2)
D^{A\,\lambda\delta}(q)\nonumber\\
\times{\cal V}_{\rho\sigma\kappa\delta}(p,q-p){\cal F}((q-p)^2)
D^{A\,\alpha\kappa}(q-p).
\end{align}
Let us adopt the entire function ${\cal F}(p^2)=\exp[-p^2/2\Lambda_G^2]$ in Euclidean momentum space, scaled by the gravitational energy scale $\Lambda_G$. We obtain
\begin{align}
\label{Ptensor}
\Pi^{\rm GA}_{\mu\nu\rho\sigma}(p)=-\kappa^2
\int\frac{d^4q\eta^{\lambda\delta}\eta^{\alpha\kappa}}{q^2(q-p)^2}{\cal V}_{\mu\nu\lambda\alpha}(p,q)\nonumber\\
\times{\cal V}_{\rho\sigma\kappa\delta}(p,q-p)\exp\biggl[-q^2/2\Lambda^2_G\biggr]
\exp\biggl[-(q-p)^2/2\Lambda^2_G\biggr].
\end{align}
As usual, we must add to (\ref{Ptensor}) the contributions from the tadpole photon-graviton diagrams and the invariant measure diagram.

We observe that from power counting of the momenta in the integral (\ref{Ptensor}), we obtain
\begin{align}
\Pi^{\rm GA}_{\mu\nu\rho\sigma}(p)\sim
\kappa^2N_{\mu\nu\rho\sigma}(\Lambda_G,p^2),
\end{align}
where $N_{\mu\nu\rho\sigma}(\Lambda_G,p^2)$ is a finite contribution to $\Pi^{\rm GA}_{\mu\nu\rho\sigma}(p)$.  ${{{\Pi^{\rm GA}_\mu}^\mu}^\sigma}_\sigma(p)$ vanishes at $p^2=0$, as it should because of gauge invariance to this order and the massless graviton.

The vector field vertex form factor, {\it when coupled to SM gauge bosons}, will have the form
\begin{equation} {\cal E}^{\rm SM}(p^2)
=\exp\biggl[-p^2/2\Lambda_{SM}^2\biggr].
\end{equation}
If we choose $\Lambda_{SM} > 1$ TeV, then we will reproduce the low energy SM experimental results and ${\cal F}^{\rm SM}(p^2)$ becomes ${\cal F}^{\rm SM}(0)=1$ on the mass shell $p^2=0$~\cite{Moffat,Moffat2}.

\section{Cosmological Constant Problem and Quantum Gravity}

The cosmological constant problem is considered to be the most severe hierarchy problem in modern physics~\cite{Weinberg,Straumann,Rugh}. Can our quantum gravity theory solve the cosmological constant problem? The cosmological constant is a non-derivative coupling in the Lagrangian density ${\cal L}_{\rm grav}$:
\begin{equation}
\label{lambda}
{\cal L}_\lambda=-\frac{4}{\kappa^2}\lambda\sqrt{-g}.
\end{equation}
In diagrammatic terms, it is a sum of zero momentum and zero temperature vacuum fluctuation loops coupled to external gravitons. The problem is to explain why the magnitude of $\lambda$ is suppressed to be zero or a very small value when compared to observation.

We define an effective cosmological constant
\begin{equation}
\lambda_{\rm eff}=\lambda_0+\lambda_{\rm vac},
\end{equation}
where $\lambda_0$ is the `bare' cosmological constant in Einstein's classical field equations,
and $\lambda_{\rm vac}$ is the contribution that arises from the vacuum density $\lambda_{\rm vac}=8\pi G_N\rho_{\rm vac}$. Already
at the SM electroweak scale $\sim 10^2$ GeV, a calculation of the vacuum density $\rho_{\rm vac}$, based on local quantum field theory results in a discrepancy with the universe's critical energy density:
\begin{equation}
\label{vacbound}
\rho_{\rm vac}\lesssim 10^{-47}\, ({\rm GeV})^4,
\end{equation}
of order $10^{55}$, resulting in a a severe fine tuning problem, since the virtual quantum fluctuations giving rise to $\lambda_{\rm vac}$ must cancel $\lambda_0$ to an unbelievable degree of accuracy. The corresponding bound on $\lambda_{\rm vac}$ is
\begin{equation}
\label{lambdabound}
\lambda_{\rm vac}\lesssim 10^{-84}\,{\rm GeV}^2.
\end{equation}
If we choose the quantum gravity scale for gravity coupled to the vacuum density, $\Lambda_{\rm Gvac}\lesssim 10^{-3}$ eV, then our quantum gravity theory leads to a suppression of the gravitational quantum corrections to $\lambda_0$ for $p^2 > \Lambda_{\rm Gvac}^2$. This suppresses $\lambda_{\rm vac}$ below the observational bound (\ref{lambdabound}).

A suppression of the vacuum energy fluctuation loops coupled to gravitons with the choice of energy scale, $\Lambda_{\rm Gvac}\lesssim 10^{-3}$ eV, suffers from a violation of the E\"otv\"os experiment~\cite{Braginsky} when a hydrogen atom is weighed in the gravitational field of the sun or Earth~\cite{Polchinski,Masso}. The vacuum polarization contribution to the Lamb shift when coupled to a graviton is known to give a non-zero contribution to the energy of the atom, and the equivalence principle requires that it couple to gravity. We know to a precision of one part in $10^6$ that the gravitational effect in the vacuum polarization contribution to the Lamb shift actually exists. The effect is larger for quark-antiquark loops than electron-positron loops.

The vacuum energy loops are coupled to external gravitons in the absence of sources as determined by (\ref{lambda}). The problem with the weighing of the hydrogen atom in a gravitational field is connected to the presence of matter sources such as a nucleus (protons). For the Lamb shift it involves the electrostatic energy of the nucleus, while for the calculation of the Casimir effect~\cite{Casimir,Milloni}, the physical system requires conducting metal plates. The vacuum loops coupled to gravitons in (\ref{lambda}) describes a physical picture different from the one consisting of vacuum loops in the presence of matter sources. The latter needs boundary conditions in space determined by the presence of matter. {\it We must understand why the zero-point energy gravitates in atomic environments and not in vacuum.} We also have to understand why this suppression of the vacuum energy coupled to gravity occurs in our vacuum and not in the vacuum associated with $SU(2)\times U(1)$ weak interactions. This problem is solved in the UV complete EW model which does not possess a Higgs mechanism vacuum state and a Higgs particle~\cite{Moffat,Moffat2}. In this model there is no symmetric $SU(2)\times U(1)$ state in which electrons, quarks and $W$ and $Z$ particles are massless. In the absence of a Higgs potential the $\rho_{\rm vac}$ produced by spontaneous EW symmetry breaking of the vacuum is zero.

A possible quantum gravity solution to the cosmological constant problem can be implemented, if we choose for the source-free, vacuum coupling of gravitons to vacuum fluctuation loops an energy scale $\Lambda_{\rm Gvac}\lesssim 10^{-3}$ eV, while for the vacuum loops in the presence of matter sources with spacetime boundary conditions, we have $\Lambda_{\rm Gmattvac}\gtrsim 1$ MeV. Thus, for the hydrogen atom and for couplings of SM particle loops to gravitons in high energy accelerators there would not be a conflict with the equivalence principle. The calculation of the zero-point vacuum density $\rho_{\rm vac}$ in space for $\Lambda_{\rm Gvac}\lesssim 10^{-3}$ eV would be sufficiently suppressed to not be in conflict with (\ref{lambdabound}). The question now arises as to how the vertex coupling gravitons to the vacuum fluctuation loops knows how to distinguish between the fluctuation loops in a vacuum and the loops in atomic and molecular systems? For the case of a pure local, point-like vertex in quantum gravity with $\sqrt{G}=\sqrt{G_N}$ there is no way to distinguish the two physical systems. However, for the case of our quantum gravity in which $\sqrt{G}=\sqrt{G_N}{\cal F}$, the nonlocal nature of the entire function ${\cal F}$ could distinguish between a pure vacuum and a material system with boundary conditions.

The UV complete quantum gravity has been applied to the problem of black holes~\cite{Nicolini}. The geometry of the black hole solution is regular. In place of the curvature singularity extreme energy fluctuations in the gravitational field at small length scales provide an effective cosmological constant in a region described by a de Sitter space. The UV complete quantum gravity removes curvature singularities in black holes.

\section{\bf Conclusions}

We have formulated a UV complete quantum gravity theory by extending the vertex coupling strength $\sqrt{G_N}$ to more general entire functions $\sqrt{G}=\sqrt{G_N}{\cal F}$ in spacetime and in momentum space. This circumvents the problem of the lack of finiteness and renormalizability of quantum gravity for $\sqrt{G}=\sqrt{G_N}$. An ongoing project is to find methods to restrict the choice of entire functions in the UV complete standard model and in our UV complete quantum gravity. It is clear that the standard choice of constant coupling strengths and local point interactions in QFT results in a restricted physical picture in which a Higgs mechanism and a physical Higgs particle are required to obtain a renormalizable EW theory. It also leads to the problem of the non-renormalizability of quantum gravity in a perturbative approach to the problem. There seems to be no obvious physical reason why nature should restrict itself to the special case of entire functions ${\cal E}=1$ and ${\cal F}=1$ in QFT. Releasing QFT from this restriction opens up the possibility of UV complete QFTs. However, experiments at the LHC should decide whether a Higgs particle exists and the high energy behavior of running coupling constants in the standard model. If the EW coupling constants $g$ and $g'$ behave as $\sim 1/\ln(s)$ for $\sqrt{s} > 1-2$ TeV where $\sqrt{s}$ is the center-of-mass energy, then this proves experimentally that at the high energies available at the LHC the coupling constants do run as predicted by standard renormalization theory. On the other hand, if the $g$ and $g'$ run with a behavior different than $1/\ln(s)$ for $\sqrt{s} > 1-2$ TeV, then this will be a clear signature that the entire functions at vertices of Feynman diagrams are not restricted to constant coupling constants in the QFT action.  Such an experimental discovery would result in a fundamental revision of QFT.

By assuming a constant energy scale $\Lambda_{\rm Gvac}\lesssim 10^{-3}$ eV for vacuum fluctuation loops coupled to gravitons in a vacuum, and the energy scale $\Lambda_{\rm Gmattvac}\gtrsim 1$ Mev for vacuum loops cb oupled to gravity in material systems, the suppression of the cosmological constant $\lambda$ can be achieved without canceling the gravitational effects in the Lamb shift in hydrogen, or physical vacuum fluctuations in atomic and molecular systems such as the Casimir effect.

\section*{Acknowledgements}

I thank Viktor Toth, Martin Green and Laurent Freidel for helpful and stimulating discussions. This work was supported by the Natural Sciences and Engineering Research Council of Canada. Research at the Perimeter Institute for Theoretical Physics is supported by the Government of Canada through NSERC and by the Province of Ontario through the Ministry of Research and Innovation (MRI).

\end{document}